\documentclass[letterpaper, 10 pt, conference]{ieeeconf} 
\IEEEoverridecommandlockouts
\usepackage{graphics} 
\usepackage{epsfig} 
\usepackage{mathptmx} 
\usepackage{times} 
\usepackage{amsmath} 
\usepackage{amssymb}  

\usepackage{subcaption}
\usepackage{xcolor}
\usepackage[cmintegrals]{newtxmath}
\usepackage{cite}
\usepackage{amsmath,amssymb,amsfonts}
\usepackage{algorithmic}
\usepackage{algorithm,algorithmic}
\usepackage{graphicx}
\usepackage{textcomp}
\usepackage{tikz}
\usepackage{xcolor}

\newcommand{\dd}{\, \mathrm{d}}
\def\BibTeX{{\rm B\kern-.05em{\sc i\kern-.025em b}\kern-.08em
		T\kern-.1667em\lower.7ex\hbox{E}\kern-.125emX}}
\usepackage{mathtools}
\newtheorem{mytheorem}{Theorem}

\newtheorem{myremark}{Remark}
\newtheorem{myassumption}{Assumption}
\newtheorem{myproblem}{Problem}

\newtheorem{myexample}{Example}

\newtheorem{mydefinition}{Definition}
\newtheorem{myproof}{proof}

\title{\LARGE \bf
Vehicular Resilient Control Strategy for a Platoon of Self-Driving Vehicles under DoS Attack 
}

\author{Hassan Mokari$^{1}$, Yufei~Tang$^{1}$
\thanks{**This work was supported in part by the U.S. National Science Foundation (NSF) through Grants CMMI-2145571 and OAC-2320972.}
\thanks{$^{1}$H. Mokari and Y. Tang are with the Department of Electrical Engineering and Computer Science, Florida Atlantic University, Boca Raton, FL 33431, USA. {\tt\small  \{hmokari2022, tangy\}@fau.edu}.}
}

\def\BibTeX{{\rm B\kern-.05em{\sc i\kern-.025em b}\kern-.08em h T\kern-.1667em\lower.7ex\hbox{E}\kern-.125emX}}

\newcommand\copyrighttext{%

  \footnotesize \textcopyright 2024 IEEE. Personal use of this material is permitted. Permission from IEEE must be obtained for all other uses, in any current or future media, including reprinting/republishing this material for advertising or promotional purposes, creating new collective works, for resale or redistribution to servers or lists, or reuse of any copyrighted component of this work in other works.
}

\newcommand\copyrightnotice{
\begin{tikzpicture}[remember picture,overlay]
\node[anchor=south,yshift=20pt] at (current page.south) {\fbox{\parbox{\dimexpr\textwidth-\fboxsep-\fboxrule\relax}{\copyrighttext}}};
\end{tikzpicture}
}

\begin{document}

\maketitle
\copyrightnotice
\thispagestyle{empty}
\pagestyle{empty}


\begin{abstract}

In a platoon, multiple autonomous vehicles engage in data exchange to navigate toward their intended destination. Within this network, a designated leader shares its status information with followers based on a predefined communication graph. However, these vehicles are susceptible to disturbances, leading to deviations from their intended routes. Denial-of-service (DoS) attacks, a significant type of cyber threat, can impact the motion of the leader. This paper addresses the destabilizing effects of DoS attacks on platoons and introduces a novel vehicular resilient control strategy to restore stability. Upon detecting and measuring a DoS attack, modeled with a time-varying delay, the proposed method initiates a process to retrieve the attacked leader. Through a newly designed switching system, the attacked leader transitions to a follower role, and a new leader is identified within a restructured platoon configuration, enabling the platoon to maintain consensus. Specifically, in the event of losing the original leader due to a DoS attack, the remaining vehicles do experience destabilization. They adapt their motions as a cohesive network through a distributed resilient controller. The effectiveness of the proposed approach is validated through an illustrative case study, showing its applicability in real-world scenarios.

\end{abstract}


\section{INTRODUCTION}
A platoon refers to a collection of intelligent vehicles that interact with one another to achieve a pre-determined destination. In general, there is a vehicle referred to as a leader, and the other members of the group tend to follow this leader \cite{1}. It is important to highlight that every follower pursues the leader at an identical speed and maintains a pre-defined inter-vehicle spacing distance \cite{2}. Communication links between vehicles (i.e., V2V) and between vehicles and infrastructure (i.e., V2I) will be formed through a defined communication topology within the platoon \cite{3++, 3++1}. To implement the intended driving objective, leader-follower consensus protocols would be utilized to obtain the original position and velocity of each agent \cite{4,5}. Due to the transmission of data via wireless communication channels between agents, vehicles in a platoon are vulnerable to cyber attacks \cite{6}. Ensuring the safety of all vehicles' overall behavior may not be entirely assured through the conventional coordinators of networks or typical approaches in information security methods \cite{pasqualetti2013attack}. 

A significant form of harmful cyber attacks is the denial-of-service (DoS) attack. This type of attack can occupy the communication channels among vehicles, causing disruptions in both incoming and outgoing data pipelines to an agent transmitted via communication links, resulting in delays \cite{7, 256+1}. Consequently, when facing a DoS attack, the targeted vehicle could be diverted from its reference route. Numerous investigations have been conducted on the consequences of DoS attacks over various applications \cite{9, 10, 11}. Nevertheless, the presence of resilient control is crucial to create a condition for retrieving the abnormal system behavior \cite{12+1}. Typical approaches involve acquiring the configuration policy for Intrusion Detection Systems from the communication layer \cite{12} and $H_{\infty}$ controllers have been proposed for dynamical systems. Research has delved into examining the stability of multi-agent systems under DoS attacks characterized by delays or packet loss. Nonetheless, the extent of investigations in this specific domain remains constrained \cite{13}. 


To counter disruptive elements like DoS attacks, adapting the communication topology to a new configuration is essential. Achieving leader-follower consensus in scenarios with switching topologies is found to be more practical, explored for both second-order and high-order systems \cite{15}. Overcoming challenges in undirected switching topologies involves a high-gain control rule \cite{17}. Directed switching topologies pose a more crucial consensus challenge, particularly studied for third-order multi-agent systems \cite{18}, \cite{19}. Additionally, Artificial Potential Field (APF) utilization has been discussed in \cite{20} to prevent collisions within a multi-agent system by generating repulsive forces. 


Additionally, studies have explored collision and obstacle avoidance for achieving specified formations within smart vehicle platoons using artificial potential fields \cite{23, 23+1}. Notably, in \cite{256}, a DoS attack is illustrated with a time-invariant delay, a representation that may significantly deviate from real-world scenarios. It's crucial to highlight that the resilient consensus algorithm in \cite{256} establishes communication with the detection and estimation algorithm only once, whereas, in practice, both algorithms should enable frequent communication for consistent updates. In addressing time-varying delays, authors in \cite{3} proposed methods to enhance system performance. However, a distributed controller to ensure a stable and secure network in case of a DoS attack on the platoon leader has not been investigated. It is evident that if the leader vehicle is under a DoS attack, all vehicles in the platoon gradually become unstable and deviate from their routes. The safety of the network is significantly jeopardized under this condition, emphasizing the need for robust vehicular resilient control.

To address the aforementioned issues, the main contributions of this paper are as follows:
\begin{enumerate}
\item We present communication graph topologies tailored to different road conditions encountered by a platoon. With the leader of the vehicle group is subjected to a DoS attack, the graph topology undergoes a transformation. The attacked vehicle, now compromised, seamlessly transitions from the role of leader to follower. While a new leader is promptly identified based on the reconfigured graph network.
\item Upon detecting a DoS attack, accomplished through the use of two incremental counters, the communication graph undergoes updates. The compromised vehicle, subjected to the attack, is systematically isolated from the other vehicles within the group based on a designed external controller. This strategic decision is driven by the rationale that segregating the attacked agent prevents the rest of the vehicles from engaging in data exchange with it to ensure that the unaffected vehicles remain unswayed from their designated routes, mitigating the impact of the attack on the overall platoon functionality.
\item To reestablish a cohesive leader-follower consensus within the platoon after an attack on a vehicle, we propose a distributed resilient controller. This controller aims to update the graph topology adaptively, isolating the affected vehicle (formerly the leader, now a follower). Through effective communication with the newly designated leader, this approach facilitates the attacked vehicle in rediscovering its correct route, ensuring a smooth transition and preserving operational efficiency within the platoon.
\end{enumerate}

The remainder of the paper is scheduled as follows: Section \ref{modeling} summarizes the modeling of the overall platoon and DoS attack. Section \ref{detection} specifies the attack detection algorithm. Section \ref{controller} presents the vehicular resilient control strategy, followed by a stability analysis in Section \ref{stability}. Section \ref{simulation} carries out simulation studies to evaluate the efficiency of the proposed approach, and Section \ref{sec:conclusion} concludes the research and discusses future work.
	
\section{MATHEMATICAL MODELLING}\label{modeling}
In this section, a dynamic model of the platoon and a distributed control approach are defined. A denial-of-service (DoS) attack is then introduced and delved into the tracking error dynamics in the platoon for subsequent analysis.

\subsection{Platoon Model}
Consider a homogeneous group of $N$ vehicles that cooperate to track and mimic the position and speed of a leader.
	
\begin{mydefinition}
	A communication graph is defined as 
	$
	\mathcal{G}=(\mathcal{V},\mathcal{E},\mathcal{A}) 
	$
	with a set of nodes 
	$
	\mathcal{V}=\{v_1,v_2,\dots,v_N\} 
	$,
	a set of edges  
	$
	\mathcal{E}\subset \mathcal{V} \times \mathcal{V} 
	$
	and an adjacency matrix 
	$
	\mathcal{A}=[a_{ij}] 
	$.
	\end{mydefinition}

In this context, a double integrator dynamical model for each vehicle is taken into consideration as:
\begin{equation}\label{dynamic_model}
    \dot s_i(t) = \zeta_i(t),~ \dot \zeta_i(t) = u_i
\end{equation}
where $s_i(t)\in\mathbb{R}$ and $\zeta_i(t)\in\mathbb{R}$ are the position and velocity of the $i$-th vehicle, respectively, and $u_i(t)$ is the control input. 
	
To achieve leader-follower consensus, consider the  distributed controller as:
	\begin{equation}\label{distributed_controller}
		u_i(t) = -\underset{j=1}{\overset{N}{\sum}} a_{ij}\Big((\hat s_i(t)-\hat s_j(t))+\gamma(\hat \zeta_i(t)-\hat \zeta_j(t))\Big)
	\end{equation}
in which, $\gamma>0$ is a design parameter to imply the closed-loop stability of the platoon \cite{12+1}, 
	$i \in \lbrace1,2,\dots,N\rbrace$
and the errors associated with both inter-vehicle spacing distance in the platoon and the velocity of each vehicle relative to the leader are:
	\begin{equation}
	    \begin{aligned}
	     &\hat s_i(t) = s_i(t) - s_N(t)-d_{i,N}\\
	     &\hat \zeta_i(t) = \zeta_i(t) - \zeta_N(t)\\
	    \end{aligned}
            \label{error-equation}
	\end{equation}
where $d_{i,N}$ is the predefined inter-vehicle spacing between the $i$-th vehicle and the leader.
	
By expanding the states of each vehicle to $X(t)=[P^T(t),V^T(t)]^T$, the closed-loop dynamics of the platoon during normal operation (without attack) can be denoted as: 
	\begin{equation}\label{theta_equation}
			\dot X(t)=\Psi X(t),\quad \Psi = \begin{bmatrix}
			0_{N \times N} & I_{N \times N}\\
			-\mathcal{L}& -\gamma \mathcal{L}
			\end{bmatrix}
	\end{equation}	
where the augmented vectors of position and velocity of all vehicles are specified as 
	$
	P(t)=[\hat s_1(t),\hat s_2(t),\dots,\hat s_N(t)]^T
	$
	and
	$
	V(t)=[\hat \zeta_1(t),\hat \zeta_2(t),\dots,\hat \zeta_N(t)]^T
	$
respectively. Furthermore, $\Psi \in \mathbb{R}^{2N \times 2N}$ and $\mathcal{L}\in \mathbb{R}^{N \times N}$ determine the Laplacian matrix of the relevant communication topology \cite{256}.
	
The closed-loop model presented in \eqref{theta_equation} will serve as the reference model for attack detection, a concept that will be elaborated upon in section \ref{detection}. 

\begin{myassumption}
In all scenarios, the communication graph of the platoon includes a spanning tree with a leader as a root node. 
\end{myassumption}

\subsection{DoS Attack Model}
A DoS attack is a type of cyber attack that temporarily disrupts the ability of vehicles to communicate and exchange information over a network. A DoS attack can be modeled as a delay in the communication link between any two distinct vehicles within the platoon. In this paper, we take into account that the time delay caused by a DoS attack is variable over time and is not known in advance. Considering a platoon comprising $N$ agents, it is assumed that vehicle $j^*$ has been subjected to an attack. The overall strategy of employing distributed control for the vehicles within the platoon is:
	\begin{equation}
		\begin{split}	
				u_i(t)&= -\Big[\sum_{j\in {\mathcal{N}_i-{j^{*}}}} a_{ij}\Big((\hat s_i(t)-\hat s_j(t))+\gamma(\hat \zeta_i(t)-\hat \zeta_j(t))\Big)\Big]\\
		&\quad - a_{i{j^{*}}}\Big((\hat s_i(t)-\hat s_{j^{*}}(t-\tau(t)))+\gamma(\hat \zeta_i(t)-\hat \zeta_{j^{*}}(t-\tau(t)))\Big)
		\\
		&\quad \pm  \Big[a_{i{j^{*}}}\Big((\hat s_i(t)-\hat s_{j^{*}}(t))+\gamma(\hat \zeta_i(t)-\hat \zeta_{j^{*}}(t))\Big)
		\Big]\\
		&=-\Big [\sum_{j\in \mathcal{N}_i} a_{ij}\Big((\hat s_i(t)-\hat s_j(t))+\gamma(\hat \zeta_i(t)-\hat \zeta_j(t))\Big)\Big]\\
		&\quad+
		a_{i{j^{*}}}\Big((\hat s_{j^{*}}(t-\tau(t))-\hat s_{j^{*}}(t))+\gamma(\hat \zeta_{j^{*}}(t-\tau(t))-\hat \zeta_{j^{*}}(t)\Big)\\
		\end{split}
	\label{20+}
		\end{equation}
in which, $\mathcal{N}_i$ denotes the neighboring set of the $i$-th agent, $\tau(t)$ is an induced time-varying delay, where the upper bound is denoted by parameter $\mathcal{U}$ as:
\begin{equation}\label{bounded_delay}
	\begin{split}
		&0<\tau(t)<\mathcal{U}, ~\ \mathcal{U}+\tau(t)<2\mathcal{U}\\
		&0<\tau(t)<\mathcal{U}<\mathcal{U}+\tau(t)<2\mathcal{U}\\
		&\dot \tau(t)<d<1 
	\end{split}
\end{equation}
where $d$ is the upper bound of the derivative of delay.
\begin{myremark}
The expression $\dot \tau(t)<d<1$ denotes a slow variation of the delay; while $1<\dot \tau(t)<d$ denotes a fast variation of delay. Here, we assume that the delay varies slowly such that $\dot \tau(t)$ is bounded.     
\end{myremark}

The following dynamic is defined as the result of \eqref{20+}:
\begin{equation}
	 \dot X(t) = \hat\Psi X(t) + \hat \Psi_1 X(t-\tau(t))
	\label{21+}
\end{equation}
where 
$
\hat \Psi \in \mathbb{R}^{2N \times 2N}
$	
denotes the impact of the state and also
$
\hat \Psi_1 \in \mathbb{R}^{2N \times 2N}
$
specifies the effect of the delayed state which are defined by:

\begin{equation}\label{}
	\begin{split}
		\begin{aligned}
	\hat \Psi = 
		\begin{bmatrix}
			0_{N \times N} & I_{N \times N}\\
			-\mathcal{\hat L}_{N \times N}& -\gamma  \mathcal{\hat L}_{N \times N}
		\end{bmatrix}
	\end{aligned}, ~\
		\begin{aligned}
		\hat \Psi_1 = 
		\begin{bmatrix}
			0_{N \times N} & 0_{N \times N}\\
			-\mathcal{G}& -\gamma  \mathcal{G}
		\end{bmatrix}
	\end{aligned}
	\end{split}
\end{equation}
		
where $\mathcal{G}=[g_{ij}]\in \mathbb{R}^{N \times N}$ is considered as
$
\mathcal{G}=\mathcal{L}-\mathcal{\hat L}
$.

We assume the leader is under DoS attack, i.e., $j^* = N$. Therefore, the platoon needs a new leader, while the attacked vehicle (former leader) will be identified as a follower of the new leader. Accordingly, all vehicles ought to track the states of the new leader labeled $N-1$ in the updated network. Therefore, the following dynamic, including the new leader and the attacked vehicle (former leader) is defined as:

\begin{equation}
	\dot X(t) = \bar\Psi X(t) + \bar\Psi_1 X(t-\tau(t))
	\label{21+}
\end{equation}
where 
$
\bar \Psi \in \mathbb{R}^{2N \times 2N}
$	
specifies the effect of the state and also
$
\bar \Psi_1 \in \mathbb{R}^{2N \times 2N}
$
determines the impact of the delayed state which are defined by:

\begin{equation}\label{}
	\begin{split}
		\begin{aligned}
	\bar \Psi = 
		\begin{bmatrix}
			0_{N \times N} & I_{N \times N}\\
			-\mathcal{\bar L}_{N \times N}& -\gamma  \mathcal{\bar L}_{N \times N}
		\end{bmatrix}
	\end{aligned}, ~\
		\begin{aligned}
		\bar \Psi_1 = 
		\begin{bmatrix}
			0_{N \times N} & 0_{N \times N}\\
			-\mathcal{\bar G}& -\gamma  \mathcal{\bar G}
		\end{bmatrix}
	\end{aligned}
	\end{split}
\end{equation}
		
where $\mathcal{\bar G}=[\bar g_{ij}]\in \mathbb{R}^{N \times N}$ is considered as
$
\mathcal{\bar G}=\mathcal{L}_1-\mathcal{\bar L}
$.

Hence, by separating the state of the leader from the dynamics of the error \eqref{21+}, it results in:	
	\begin{equation}\label{Z_dynamic}
		\dot \Theta(t) = \tilde \Psi \Theta(t) + \tilde \Psi_1 \Theta(t-\tau(t))
	\end{equation}
where $\Theta(t)=[P^T_z(t), V^T_z(t)]^T $ is the augmented vector of error such that 
	$
	P_z(t)=[\hat s_1(t),\hat s_2(t),\dots,\hat s_{N-1}(t)]^T
	$
and
	$
	V_z(t)=[\hat \zeta_1(t),\hat \zeta_2(t),\dots,\hat \zeta_{N-1}(t)]^T
	$.

The impact of the error signal in \eqref{Z_dynamic} is denoted by $\tilde \Psi \in \mathbb{R}^{2(N-1) \times 2(N-1)}$, which is defined as:
 \begin{equation}
		\begin{aligned}
			\tilde \Psi = 
			\begin{bmatrix}
				0_{(N-1) \times (N-1)} & I_{(N-1) \times (N-1)}\\
				-\mathcal{Q}& -\gamma \mathcal{Q}
			\end{bmatrix}
		\end{aligned}
		\label{Theta}
	\end{equation}
where $\mathcal{Q}=[q_{ij}] \in \mathbb{R}^{(N-1) \times (N-1)}$ can be calculated as
$q_{ij}={\hat l}_{ij}-{\hat l}_{Nj}$. Furthermore, $\tilde \Psi_1 \in \mathbb{R}^{2(N-1) \times 2(N-1)}$ corresponds to the effect of delayed error in the dynamic model as:
\begin{equation}
	\begin{aligned}
		\tilde \Psi_1 = 
		\begin{bmatrix}
			0_{(N-1) \times (N-1)} & I_{(N-1) \times (N-1)}\\
			-\mathcal{W}& -\gamma \mathcal{W}
		\end{bmatrix}
	\end{aligned}
\end{equation}
	where
    $\mathcal{W}=[w_{ij}] \in \mathbb{R}^{(N-1) \times (N-1)}$
    can be obtained as	
    $w_{ij}=g_{ij}-g_{Nj}$
     \cite{3}.
	
\begin{myremark}
The matrix $\mathcal{Q}$ contains all eigenvalues of the Laplacian matrix except zero, regarding to \eqref{theta_equation}.
\end{myremark}
	
\section{ATTACK DETECTION}\label{detection}
In this section, we use two incremental timers. i.e., $\mathcal{T}_{1}$ and $\mathcal{T}_{2}$, through Algorithm 1 to identify diverted vehicles under the DoS attack, which is the first step to developing a resilient controller.

\begin{algorithm}[t]
	\begin{algorithmic}[1]
		\renewcommand{\algorithmicrequire}{\textbf{Input:}}
		\renewcommand{\algorithmicensure}{\textbf{Output:}}
		\REQUIRE $ x_i(t)=[s_i(t),\zeta_i(t)]^T$, $\mathcal{T}_{i,1}=\mathcal{T}_{i,2}=0$
		\ENSURE  $\tau_i(t):=$ Measurement of delay injected by DoS
		\FOR{$i=1,2,...,N-1$}
		\STATE Receive $ x_i(t)$ and $x_{i,ref}(t)$
		\IF{$\lVert  x_i(t)-x_{i,ref}(t)\rVert \leq \varepsilon  $}
        \STATE $\tau_i(t)=0$; and continue
		\ENDIF
		\WHILE{$\lVert  x_i(t)-x_{i,ref}(t)\rVert > \varepsilon $ }
        \STATE  Vehicle $i$ is isolated; and timer $1$ is ceased
		\STATE 	$X_{i,r}=x_{i,ref}(t)$
		\STATE	DoS attack is launched into $i$-th vehicle
		{\WHILE{$(1)$}
		\IF{$ x_{i}(t)=X_{i,r}$}
		\STATE  Timer $2$ would be stopped
		\STATE  \textbf{break}
		\ENDIF
		\ENDWHILE
		\STATE $\tau_i(t) = \mathcal{T}_{i,2}-\mathcal{T}_{i,1}$; and timer 1 and timer 2 get reset}
		\ENDWHILE
		\ENDFOR
	\end{algorithmic}
	\caption{Detection and Measurement of Delay}
	\label{al1}
\end{algorithm}

The identification of a DoS attack and the measurement of the associated injected delay ($\tau_i(t)$) on the $i$-th vehicle are investigated through Algorithm \ref{al1}, utilizing the reference models \eqref{theta_equation} and \eqref{21+} \cite{3}.

Referring to Algorithm \ref{al1}, the detection of DoS attacks is fulfilled using the integration of two incremental counters for each vehicle: counter $1$ and counter $2$. Initially, both timers commence simultaneously, thus maintaining identical values in the absence of a DoS attack, as indicated in line~$4$ of the algorithm. Once a vehicle is misled because of a DoS attack, its states diverge from the states of the reference model, as referred to in line $6$. At this step, counter $1$ halts, while counter $2$ continues to tally until the state of the attacked vehicle equals the last received value, stored in the parameter $X_{i,r}$ in line $11$. Promptly thereafter, counter $2$ is also halted. Consequently, the delay is measured by the difference between the values of the two counters.

\section{RESILIENT CONTROL}\label{controller}
Here, a resilient controller designed to retrieve the integrity of the consensus system is presented within a platoon following the detection of a DoS attack. We further investigate the switching network topology transitions under the attack. Ultimately, all platoon members will converge toward a common destination outlined by the reference model.


Consider the closed-loop dynamics described in \eqref{theta_equation} and \eqref{21+} in the context of a consensus system involving $N$ vehicles. In the event of a DoS attack affecting the $j^*$-th agent, we employ Algorithm 2 to guide the compromised agent back to the leader-follower consensus state \cite{256}. This study specifically addresses scenarios where the leader vehicle is the target of a DoS attack.

When the leader vehicle experiences a DoS attack, it loses its leadership role within the platoon. Instead, it transitions to a follower role in the updated communication graph among the vehicles. Furthermore, a new vehicle positioned closest to the attacked leader is identified as the root node in the modified spanning tree and assumes the new leadership role within the platoon. The selection of a new leader requires careful consideration to minimize alterations in the communication links within the graph.

\begin{algorithm}[t]
	\caption{Resilient Consensus}
	\label{al2}
	\begin{algorithmic}[1]
		\renewcommand{\algorithmicrequire}{\textbf{Input:}}
		\renewcommand{\algorithmicensure}{\textbf{Output:}}
		\REQUIRE $x_i(t)=[s_i(t),\zeta_i(t)]^T$, $\mathcal{C}$ = Critical area lower bound  
		\ENSURE  $\zeta_i(t)$ 
		\WHILE{(1)}
		\FOR{$i=1,2,...,N-1$}
		\IF{$\tau_i(t) < \mathcal{C} $}
		\STATE Switching signal is enabled for $i$-th agent
		\ELSE
		\STATE  $\zeta_i(t)\rightarrow 0$ 
		\ENDIF
		\ENDFOR
		\ENDWHILE
	\end{algorithmic} 
\end{algorithm}

\begin{myremark}
Practically, it should be followed to gradually reduce the velocity of a vehicle when coming to a stop. Similarly, when detecting the presence of a DoS attack and observing that the measured delay exceeds the critical value ($\mathcal{C}$) for the affected vehicle, it is imperative to first minimize the speed variations of the vehicle ($\dot\zeta_i(t)< 0$), and then ensure its velocity converges to zero ($\zeta_i(t)\rightarrow0$). Following Algorithm $1$, the identification of the attacked agent and its corresponding vehicle, along with the measurement of the delay amount, can be achieved by employing the two incremental counters as defined. Subsequently, Algorithm $2$ is executed to address the affected agent.
\end{myremark} 

A vehicular resilient control strategy is used based on switching systems, enabling the reclamation of diverted vehicles under a critical condition ($\tau(t) < \mathcal{C}$) imposed by DoS attacks, where $\mathcal{C}$ denotes a design parameter bounded by the maximum delay $\mathcal{U}$ specified in \eqref{bounded_delay}. A pivotal region is dynamically adjusted based on the network's sensitivity to treat or cease the affected vehicle.
As indicated in line $3$ in Algorithm 2, when the observed delay for an agent falls below the lower threshold of the pivotal region, the switching system is leveraged to retrieve this agent to the leader-follower consensus \cite{256}. 

\begin{mydefinition}[\cite{26}]
Consider a platoon of smart vehicles with homogeneous dynamics under a distributed controller formulated as \eqref{distributed_controller} in which each agent is modeled with \eqref{dynamic_model}. If the switching signal is designed with three levels with respect to the following switching law:
\begin{equation}	
		N_{\sigma_i}(t_k,t_{k+1}) \le N_{0i}+\frac{T_i(t_k,t_{k+1})}{\tau_{ai}} 
		\label{su3}
\end{equation}
where
	$
	\tau_{ai}
	$
is introduced as mode-dependent average dwell time (MDADT),
	$
	T_i(t_k,t_{k+1})
	$
denotes the time window length that is
	$
	[t_k,t_{k+1})
	$,
	$
	N_{0i}
	$
is chatter bound and
	$
	N_{\sigma_i}(t_k,t_{k+1})
	$
specifies the number of switches in the time duration, then the switching system operates to make the attacked vehicle move correctly, and return it to the leader-follower consensus.
\end{mydefinition}

\begin{myremark}	
The switching system is designed according to \eqref{su3}, which works based on the MDADT ($\tau_{a}$). This choice ensures the stability of the platoon's consensus throughout its operation. Specifically, intervals between switches should carefully be chosen to guarantee overall stability over time. In other words, the design aims to swiftly retrieve any disrupted agent to the leader-follower consensus, ensuring exponential convergence over time. For instance, when employing MDADT for stability proof, it assumes that the entire duration comprises intervals where the derivative of the Lyapunov-Krasovskii function must be negative definite in each time interval to maintain stability.		
\end{myremark}

We consider the switching rule outlined in \eqref{su3}. Conversely, if the measured delay surpasses a critical threshold, the attacked vehicle will stop, as indicated in line $6$ in Algorithm~\ref{al2}.


\section{STABILITY ANALYSIS}\label{stability}
In this section, regarding \cite{3} by introducing the augmented Lyapunov-Krasovskii function, the stability of \eqref{Z_dynamic} would be explored in the presence of a DoS attack. 

\begin{mytheorem}
The system \eqref{Z_dynamic} under \eqref{bounded_delay} would be asymptotically stable if and only if:
	\[
	\begin{bmatrix}	
		\theta_{11} & \theta_{12} & Q \\
		\theta_{21} & \theta_{22} & 0\\
		* & * & -Q
	\end{bmatrix} \prec 0
	\]
where
	\begin{flalign}	
		\theta_{11}&=\mathcal{U}^2\tilde\Psi^TQ\tilde\Psi-Q+S+\tilde\Psi^T \mathcal{H}_p + \mathcal{H}_p \tilde\Psi &\\
		\theta_{12}&=\mathcal{U}^2\tilde\Psi^TQ\tilde\Psi_1 +\mathcal{H}_p\tilde\Psi_1&\\
		\theta_{21}&=\mathcal{U}^2\tilde\Psi^T_1Q\tilde\Psi +\tilde\Psi^T_1 \mathcal{H}_p&\\	
		\theta_{22}&=\mathcal{U}^2\tilde\Psi^T_1Q\tilde\Psi_1-(1-d)S	
	\end{flalign}
\end{mytheorem}
in which, 
$
Q \in \mathbb{R}^{2(N-1) \times 2(N-1)}
$,	
$
S \in \mathbb{R}^{2(N-1) \times 2(N-1)}
$	
and 
$
\mathcal{H}_p \in \mathbb{R}^{2(N-1) \times 2(N-1)}
$	
are positive definite matrices. As \eqref{bounded_delay} denotes, $\mathcal{U}$ and $d$ are defined as the upper bound of delay and the upper bound of delay derivative, respectively.

\begin{myproof}
	Consider the following Lyapunov-Krasovskii function which contains three parts:
	\begin{equation}
		\begin{split}	
			V(t,\Theta(t),\dot \Theta(t))&=\mathcal{U}\int_{-\mathcal{U}}^{0}\int_{t+\beta}^{t}\dot
			\Theta^T(s)Q\dot \Theta(s) \dd s \dd \beta\\
			\\
			& \quad +\int_{t-\tau(t)}^{t} \Theta^T(s)S\Theta(s)\dd s +\Theta^T(t)\mathcal{H}_p\Theta(t)
		\end{split}
		\label{V}
	\end{equation}	
where the first part includes the upper bound of the delay and the second part contains the delay. However, the delay is not indicated in the third part which is associated with the switched subsystem as we must imply the stability of the proposed resilient control strategy.

In the following, we study these three parts individually. The first part is denoted as:
	\begin{equation}\label{V1}
		V_1(t,\Theta(t),\dot \Theta(t))=\mathcal{U}\int_{-\mathcal{U}}^{0}\int_{t+\beta}^{t}\dot \Theta^T(s)Q\dot \Theta(s) \dd s \dd \beta
	\end{equation}
	\begin{equation}	
		V_1(t,\Theta(t),\dot \Theta(t))=\mathcal{U}\int_{t-\mathcal{U}}^{t}\int_{-\mathcal{U}}^{s-t}\dot \Theta^T(s)Q\dot \Theta(s) \dd \beta \dd s  
	\end{equation}
	\begin{equation}
		V_1(t,\Theta(t),\dot \Theta(t))=\mathcal{U} \int_{t-\mathcal{U}}^{t} (s-t+\mathcal{U}) \dot \Theta^T(s)Q \dot \Theta(s) \dd s 
	\end{equation}	
	then, by differentiating $V_1(t,\Theta(t),\dot \Theta(t))$, we achieve:
	\begin{equation}
		\dot V_1= \mathcal{U}^2 \dot \Theta^T(t)Q\dot \Theta(t)-\mathcal{U}\int_{t-\mathcal{U}}^{t} \dot \Theta^T(s)Q \dot \Theta(s) \dd s 
		\label{eqq2}
	\end{equation}
	\begin{equation}\label{eq2}
		\dot V_1=
		\begin{cases}
			\dot V_{11}=\mathcal{U}^2 \dot \Theta^T(t)Q\dot \Theta(t)\\
			\dot V_{12}=-\mathcal{U}\int_{t-\mathcal{U}}^{t} \dot \Theta^T(s)Q \dot \Theta(s) \dd s	
		\end{cases}
	\end{equation}
	by incorporating \eqref{Z_dynamic} in $\dot V_{11}$, it yields to:	
	
	\begin{equation}
		\dot V_{11}=\mathcal{U}^2(\Theta^T(t)\tilde \Psi+\Theta^T(t-\tau(t))\tilde \Psi^T_1)Q(\tilde\Psi \Theta(t)+\tilde\Psi_1 \Theta(t-\tau(t)))
		\label{q*}
	\end{equation}	
 
To obtain linear matrix inequalities (LMIs), we define an augmented vector of errors as: 	
	\begin{equation}\label{eq3}
		\xi(t) =
		\begin{bmatrix}
			\Theta^T(t) & \Theta^T(t-\tau(t)) & \Theta^T(t-\mathcal{U}) 
		\end{bmatrix}^T
	\end{equation}
and \eqref{q*} can be rewritten as:	
	\begin{equation}
		\begin{split}
			\dot V_{11}&= (\xi^T
			\begin{bmatrix}
				\mathcal{U}\tilde\Psi^T Q&
				\mathcal{U}\tilde\Psi^T_1 Q&
				0
			\end{bmatrix}^T
			)Q^{-1}(
			\begin{bmatrix}
				\mathcal{U} Q\tilde\Psi  &
				\mathcal{U} Q\tilde\Psi_1 &
				0
			\end{bmatrix}
			\xi)\\
			& = \xi^T(t) 
			\begin{bmatrix}
				\mathcal{U}\tilde\Psi^T Q^{\frac12}\\
				\mathcal{U}\tilde\Psi^T_1 Q^{\frac12}\\
				0
			\end{bmatrix}
			\begin{bmatrix}
				\mathcal{U} Q^{\frac12}\tilde\Psi  &
				\mathcal{U} Q^{\frac12}\tilde\Psi_1 &
				0
			\end{bmatrix}
			\xi(t)
		\end{split}
	\end{equation}	
where $0 \in \mathbb{R}^{2(n-1) \times 2(n-1)}$ is a zero matrix, and we can re-formulate it as:
	\begin{equation}\label{Vd11}
		\dot V_{11}=\xi^T(t)
		\begin{bmatrix}
			\mathcal{U}^2\tilde\Psi^TQ\tilde\Psi & \mathcal{U}^2\tilde\Psi^TQ\tilde\Psi_1 & 0 \\
			\mathcal{U}^2\tilde\Psi^T_1Q\tilde\Psi & \mathcal{U}^2\tilde\Psi^T_1Q\tilde\Psi_1 & 0\\
			0 & 0 & 0
		\end{bmatrix}
		\xi(t) 
	\end{equation}

\begin{figure*}[t]
    \centering
    \includegraphics[width=0.9\linewidth]{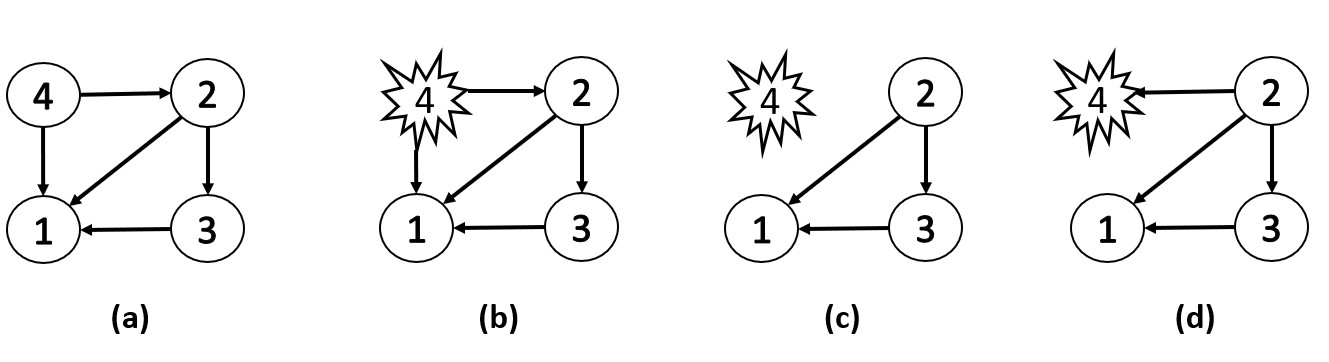}
    \caption{Communication network diagrams with 4 vehicles: (a) Represents the normal mode (no attack); (b) Implies the leader is under DoS attack and this attached is detected; (c) Illustrates the attacked vehicle has been isolated; (d) Signifies vehicle 2 becomes the new leader and communicates with the platoon, aiming to reestablish a new leader-follower consensus.}
    \label{graph}
\end{figure*}
 
Regarding to \eqref{eq2}, by taking derivation of the second part of \eqref{eqq2}, it yields:
	\begin{equation}
		\dot V_{12}=-\mathcal{U}\int_{t-\mathcal{U}}^{t} \dot \Theta^T(s)Q \dot \Theta(s)\dd s
		\label{eqq3}
	\end{equation}
and then, using Jensen's inequality \cite{278}, \eqref{eqq3} can be simplified as:  	
	\begin{equation}
		\dot V_{12}
		\le -(\int_{t-\mathcal{U}}^{t} \dot \Theta^T(s) \dd s)Q(\int_{t-\mathcal{U}}^{t} \dot \Theta(s) \dd s)
	\end{equation}
 \begin{equation}
		\dot V_{12}\le (-\Theta^T(t)+ \Theta^T(t-\mathcal{U}))Q(\Theta(t)- \Theta(t-\mathcal{U}))
	\end{equation}
	\begin{equation}
		\dot V_{12}\le\xi^T(t)
		\begin{bmatrix}
			-Q & 0 & Q\\
			0 & 0 & 0\\
			Q & 0 & -Q		
		\end{bmatrix}
		\xi(t)
		\label{Vd12}
	\end{equation}		

Regarding to \eqref{Vd11} and \eqref{Vd12}, the LMI corresponding to the first part in Lyapunov function \eqref{V1} is:
	\begin{equation}\label{vd1}
		\dot V_1\le\xi^T(t)
		\begin{bmatrix}	
			\psi_{11} & \psi_{12} & Q \\
			\psi_{21} & \psi_{22} & 0\\
			* & * & -Q
		\end{bmatrix}
		\xi(t)
	\end{equation}
	where
\begin{equation*}
    \psi_{11}\&=\mathcal{U}^2\tilde\Psi^TQ\tilde\Psi-Q; ~\ \psi_{12}\&=\mathcal{U}^2\tilde\Psi^TQ\tilde\Psi_1	
\end{equation*}
\begin{equation*}
    \psi_{21}\&=\mathcal{U}^2\tilde\Psi^T_1Q\tilde\Psi; ~\ \psi_{22}\&=\mathcal{U}^2\tilde\Psi^T_1Q\tilde\Psi_1
\end{equation*}  
 
Afterward, the second term of \eqref{V} can be written as:	
	\begin{equation}\label{V2}	
		V_2(t,\Theta(t))=\int_{t-\tau(t)}^{t} \Theta^T(s)S\Theta(s)\dd s
	\end{equation}
and by differentiating the above function, it yields:	
	\begin{equation}			
		\dot V_2=\Theta^T(t)S\Theta(t)-(1-\dot \tau(t))\Theta^T(t-\tau(t))S\Theta(t-\tau(t))
	\end{equation}
	
 Since, 
	$
	\dot \tau(t)<d
	$
	is specified with respect to \eqref{bounded_delay}, the following equation is obtained:	
	\begin{equation}	
		\dot V_2 \le \Theta^T(t)S\Theta(t)-(1-d)\Theta^T(t-\tau(t))S\Theta(t-\tau(t))
	\end{equation}
and by incorporating \eqref{eq3}, the previous inequality can be rewritten as the following LMI:
	\begin{equation}
		\dot V_2\le			
		\xi^T(t)
		\begin{bmatrix}
			S&0&0\\0&-(1-d)S&0\\0&0&0
		\end{bmatrix}
		\xi(t)
		\label{vd2}
	\end{equation}	
thus the third term of \eqref{V} is:
	\begin{equation}\label{third_term}
		V_3(t,\Theta(t))=\Theta^T(t)\mathcal{H}_p\Theta(t)
	\end{equation}
in which, $\mathcal{H}_p$ is a positive definite matrix, 
	switching signal would be also specified as $\sigma(t)$ which is in the finite set $K=\lbrace 1,2,\dots,S \rbrace$, while $S$ is defined as the number of subsystems and $\forall (\sigma(t_i)=p, \sigma(t_i^-)=f) \in K \times K$, $p \neq f$. By taking derivative of \eqref{third_term}, and using \eqref{Z_dynamic}, we achieve:
	\begin{equation}
      \begin{split}
       \dot V_3&= \text{sym}(\dot \Theta^T(t)\mathcal{H}_p\Theta(t))\\
       &= \dot \Theta^T(t)\mathcal{H}_p\Theta(t)+\Theta^T(t)\mathcal{H}_p\dot \Theta(t)\\
       &= (\Theta^T(t)\tilde \Psi^T + \Theta^T(t-\tau(t))\tilde \Psi^T_{1})\mathcal{H}_p \Theta(t)\\
       &\quad + \Theta^T(t) \mathcal{H}_p (\tilde \Psi \Theta(t) + \tilde \Psi_1 \Theta(t-\tau(t)))\\
       &= \Theta^T(t)(\tilde\Psi^T \mathcal{H}_p + \mathcal{H}_p \tilde\Psi)\Theta(t)\\
       &\quad + \Theta^T(t)\mathcal{H}_p\tilde\Psi + \Theta^T(t-\tau(t))\tilde\Psi^T_1 \mathcal{H}_p \Theta(t)
        \end{split}
	\end{equation}
	where based on \eqref{su3}, if we write the previous simplified function as \eqref{vd1} and \eqref{vd2}, its LMI is presented as:		
	\begin{equation}
		\dot V_3=\xi^T(t)
		\begin{bmatrix}
			\tilde\Psi^T \mathcal{H}_p + \mathcal{H}_p \tilde\Psi&\mathcal{H}_p\tilde\Psi_1 &0\\ \tilde\Psi^T_1 \mathcal{H}_p&0&0\\0&0&0
		\end{bmatrix}
		\xi(t)
		\label{vdd3}
	\end{equation}

Finally, regarding to \eqref{vd1}, \eqref{vd2} and \eqref{vdd3}, the total LMI corresponding to \eqref{V} will be written as:
	\begin{equation}
		\begin{split}
			\dot V&=\dot V_1
			+\dot V_2+\dot V_3\\& \le \xi^T(t)		
			\begin{bmatrix}	
				\theta_{11} & \theta_{12} & Q \\
				\theta_{21} & \theta_{22} & 0\\
				* & * & -Q
			\end{bmatrix} \xi(t)
		\end{split}
		\label{kol}
	\end{equation}
	where
	\begin{flalign*}	
		\theta_{11}&=\mathcal{U}^2\tilde\Psi^TQ\tilde\Psi-Q+S+\tilde\Psi^T \mathcal{H}_p + \mathcal{H}_p \tilde\Psi &\\
		\theta_{12}&=\mathcal{U}^2\tilde\Psi^TQ\tilde\Psi_1 +\mathcal{H}_p\tilde\Psi_1&\\
		\theta_{21}&=\mathcal{U}^2\tilde\Psi^T_1Q\tilde\Psi +\tilde\Psi^T_1 \mathcal{H}_p&\\	
		\theta_{22}&=\mathcal{U}^2\tilde\Psi^T_1Q\tilde\Psi_1-(1-d)S
	\end{flalign*}

Now, according to \eqref{kol}, if the following matrix would be negative definite, it yields the asymptotic stability of \eqref{Z_dynamic}, and this completes the proof.
\begin{equation}	
	\begin{bmatrix}		
		\theta_{11} & \theta_{12} & Q \\
		\theta_{21} & \theta_{22} & 0\\
		* & * & -Q
	\end{bmatrix}\prec 0
\end{equation}
\end{myproof}

\begin{myremark}
It is important to acknowledge that the incremental counters used to estimate delay should be reset to release the estimated delay amount such that the proposed control algorithm can be used in different scenarios. 
\end{myremark}

\section{SIMULATION RESULTS}\label{simulation}
In this section, we validate our proposed resilient control strategy through a simulation study of a homogeneous platoon comprising $4$ vehicles.

Consider the following dynamic model of each vehicle in the platoon as described in \eqref{dynamic_model}, incorporating a stabilizing local feedback controller:
			\begin{equation}
		\dot x_i(t)=
		\begin{bmatrix}
		0&1\\
		-7&-6	
		\end{bmatrix}x_i(t)+
		\begin{bmatrix}
		0\\1
		\end{bmatrix}\hat u_i(t)	
		\end{equation}

\begin{figure}[t]
    \centering
    \begin{subfigure}[b]{0.5\textwidth}
        \includegraphics[width=\textwidth]{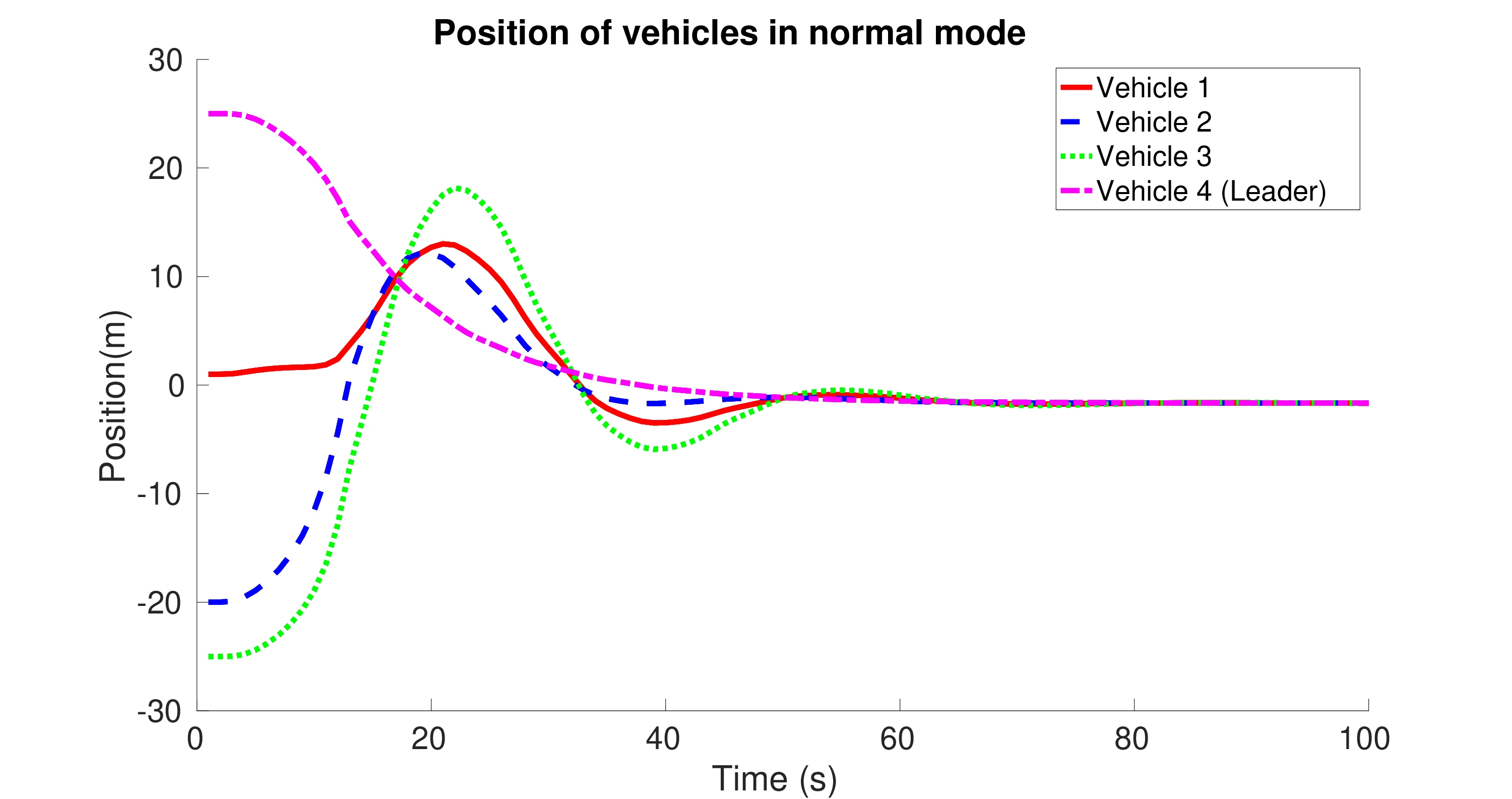}
        \label{x3-normal}
    \end{subfigure}
    \begin{subfigure}[b]{0.5\textwidth}
        \includegraphics[width=\textwidth]{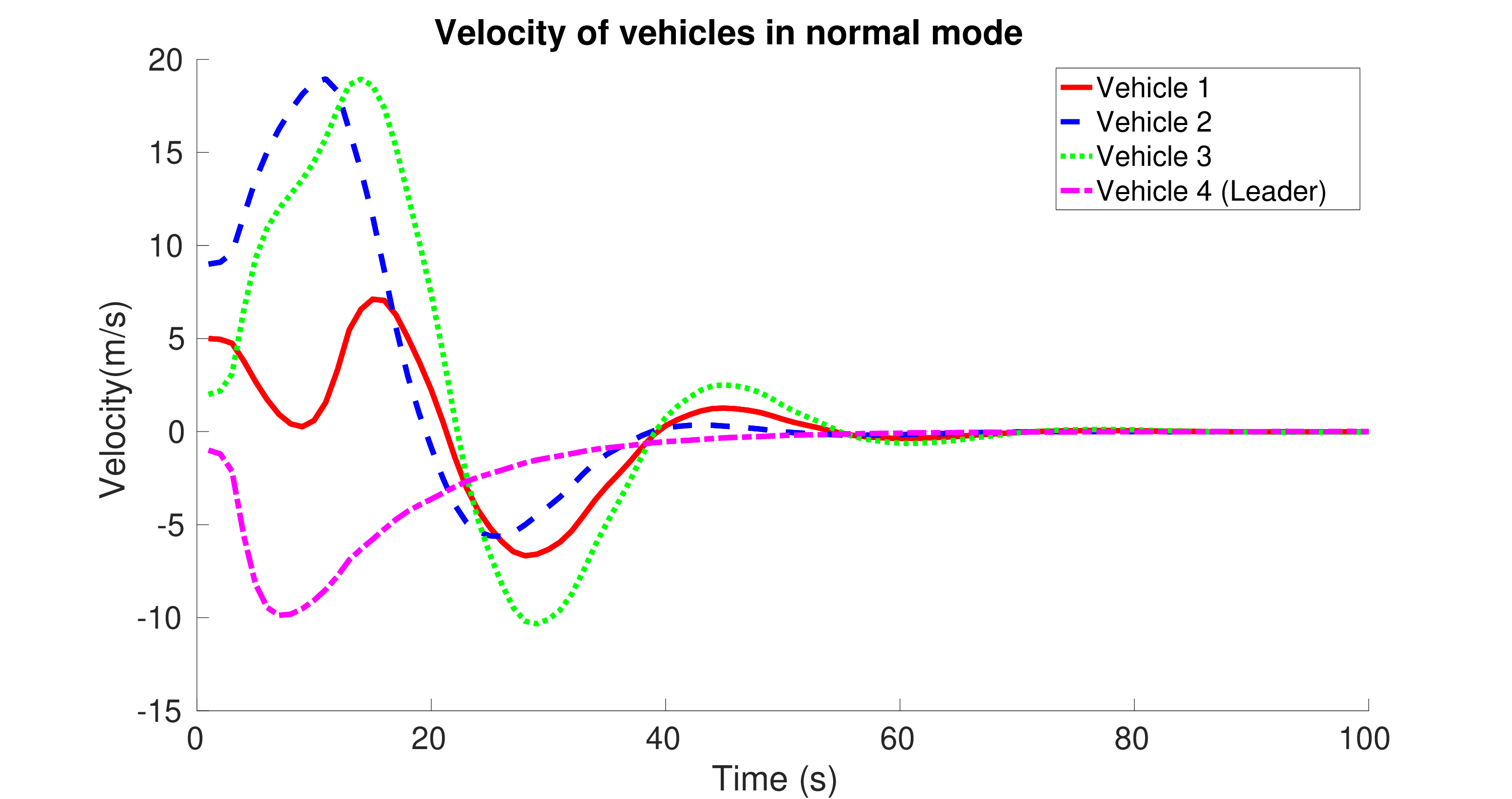}
        \label{v3-normal}
    \end{subfigure}
    \caption{(Top) Position of vehicles and (Bottom) Velocity of vehicles without attacks. Vehicle 4 is the leader.}
    \label{fig:normal}
\end{figure}

In the communication structure illustrated in Fig.~\ref{graph}(a), the $4$-th agent is identified as the leader within the consensus system of the platoon. Fig.~\ref{fig:normal} depicts the position and velocity of vehicles in the normal mode (no attack).

We hypothesize that the $4$-th vehicle, acting as the leader of the platoon, is under a DoS attack as depicted in Fig.~\ref{graph}(b). In the initial step, the vehicle under a DoS attack is identified using two incremental counters according to Algorithm $1$, where the delay amount is measured. In the second step, the attacked vehicle will no longer play the role of leader in the platoon, while it will transition to a follower in the new communication graph for other vehicles. The $2$-th vehicle which has the closest position to the $4$-th vehicle will be denoted as a root node in the updated spanning tree (shown in Fig.~\ref{graph}(c)) and will play a new leader role in the platoon. Choosing a new leader for the platoon requires careful consideration to ensure that the selection of a vehicle minimally alters the communication links within the graph. 

\begin{figure}[t]
    \centering
    \includegraphics[width=1\linewidth]{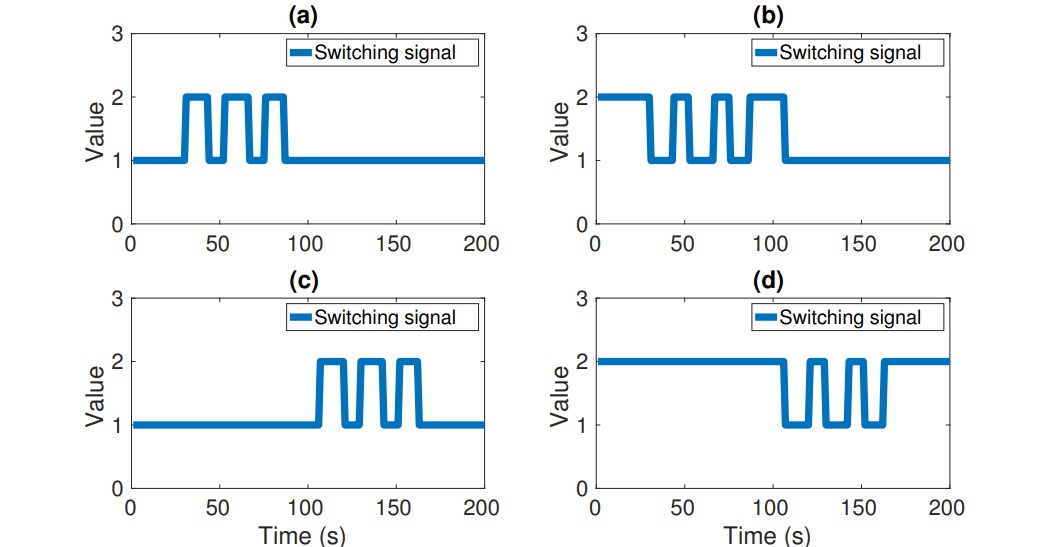}
    \caption{Switching signals.}
    \label{switching}
\end{figure}

In the third step, the attacked vehicle (previous leader) is now identified as a follower vehicle by applying Algorithm $2$. The detected attacked agent is stopped or returned to leader-follower consensus according to the delay amount measured by Algorithm $1$. This system operates based on the switching signal depicted in Fig.~\ref{switching}. In this case study, we assume that the lower bound for the critical area mentioned in Algorithm \ref{al1} is 15 seconds. Regarding \eqref{theta_equation}, \eqref{bounded_delay}, and \eqref{Z_dynamic}, in the absence of a DoS attack, consensus among the leader and followers within the platoon would be attained. In Fig. \ref{graph}, we depict the corresponding graph topology. 

\begin{figure}[t]
    \centering
    \includegraphics[width=1\linewidth]{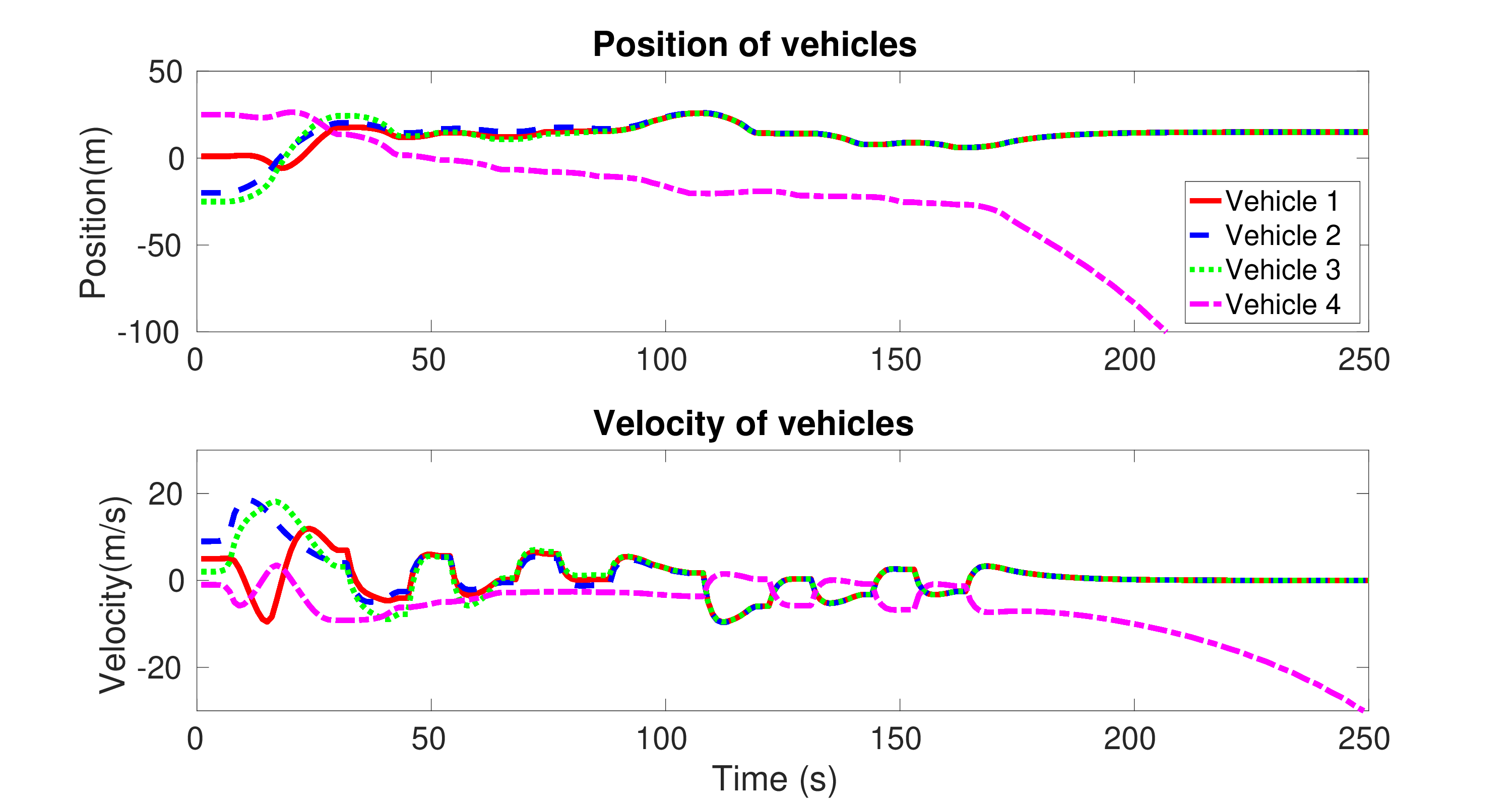}
    \caption{Position and velocity of vehicles under DoS attack. Vehicle 4 is under attack and is detected and isolated.}
    \label{xv3-isounderattack}
\end{figure}

When the $4$-th vehicle is detected as an agent under DoS attack, the amount of delay is calculated. If the value of delay is less than $15$s, according to the first step, this vehicle must be isolated from the rest of the agents in the platoon based on two switching signals depicted in Fig.~\ref{switching}(a) and Fig.~\ref{switching}(b). Fig.~\ref{xv3-isounderattack} depicts the position and velocity of each vehicle in this group.

\begin{figure}[t]
    \centering
    \includegraphics[width=1\linewidth]{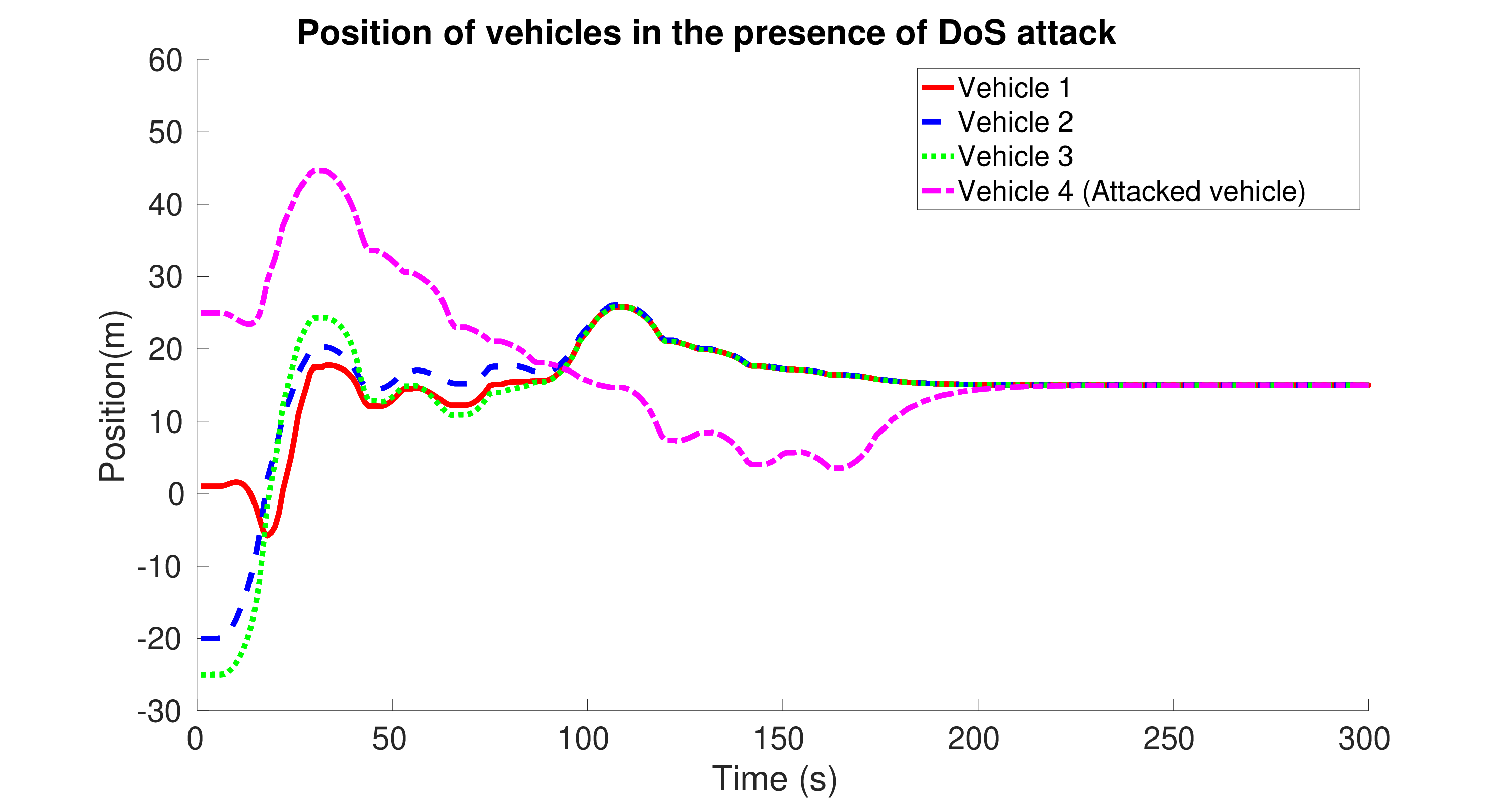}
    \caption{Position of vehicles. Vehicle 4 is under attack. Vehicle 2 is the new leader.}
    \label{x5-underattack}
\end{figure}

\begin{figure}[t]
    \centering
    \begin{subfigure}[b]{0.5\textwidth}
        \includegraphics[width=\textwidth]{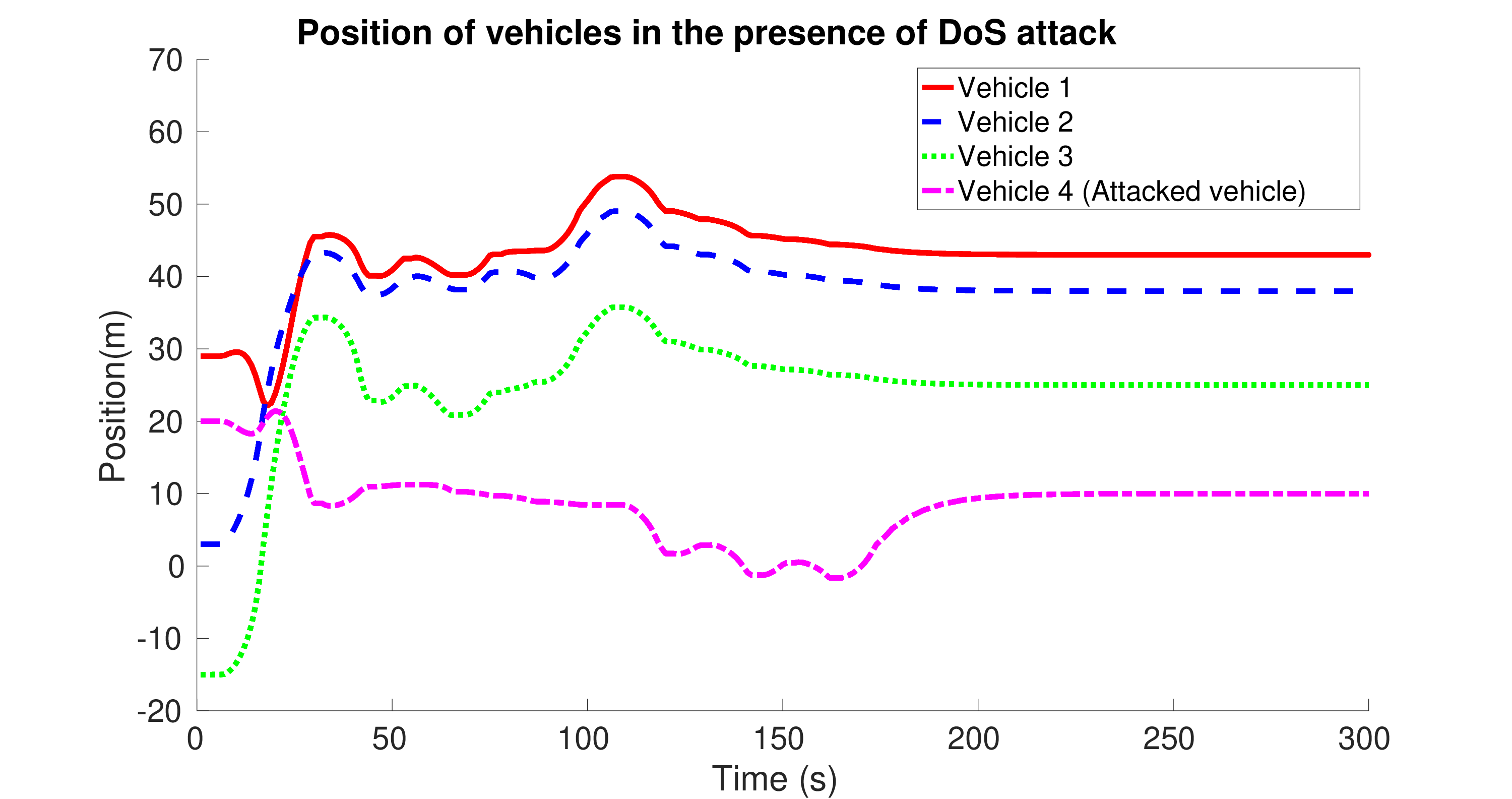}
        \label{x5-2-distunderattack}
    \end{subfigure}
    \begin{subfigure}[b]{0.5\textwidth}
        \includegraphics[width=\textwidth]{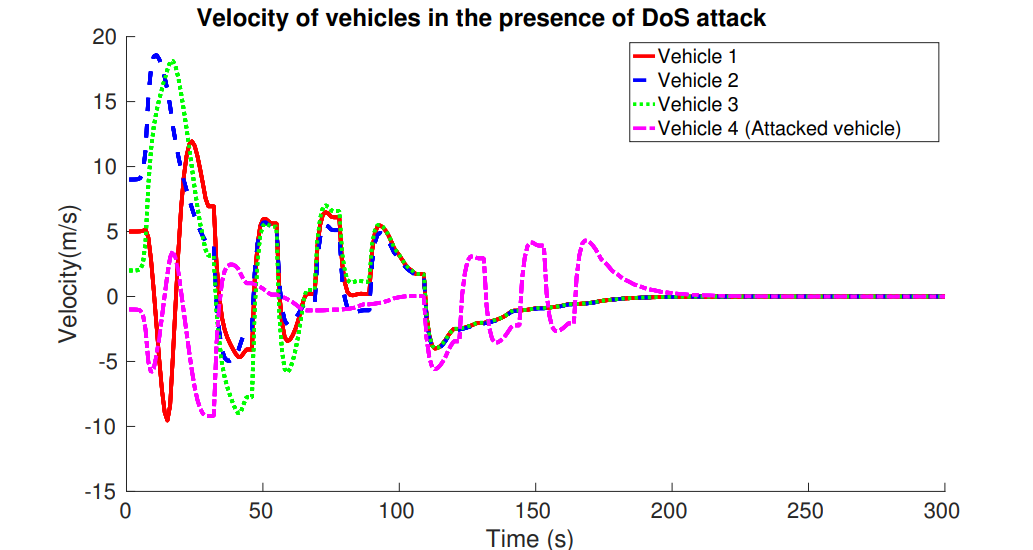}
        \label{v5-underattack}
    \end{subfigure}
    \caption{(Top) Position of vehicles and (Bottom) Velocity of vehicles under DoS attacks, regarding to the inter-vehicle spacing distance. Vehicle 2 is the leader.}
    \label{fig:normal2}
\end{figure}


The communication graph involving the isolated vehicle $4$ can be inferred from the illustration in Fig.~\ref{graph}(c). After the identification of the $4$-th vehicle (previous leader) experiencing a DoS attack at time $t=30s$ through the utilization of Algorithm $1$, the resilient control strategy defined in Algorithm $2$ is activated at that moment. Because the injected time-varying delay's value is subject to change at any given moment, it is essential to measure its quantity in a way that continuously updates the measured delay amount to reflect the current value. If the delay amount falls below the threshold of the unauthorized zone ($15$ seconds), the resilient control strategy outlined in Algorithm $2$ will be activated to retrieve the attacked vehicle by switching between two switching signals demonstrated as Fig.~\ref{switching}(c) and Fig.~\ref{switching}(d).

Following this instance, the attacked vehicle would be treated, completely. While it can receive data from adjacent vehicles, it is not permitted to transmit its information to them (shown in Fig.~\ref{graph}(d)). Likewise, the attacked agent can follow the leader's state to reach the leader-follower consensus. Therefore, the agent that has undergone treatment exhibits regular behavior, and the switching system becomes inactive under these conditions (shown in Fig.~\ref{x5-underattack}). Also, the position and velocity of the vehicles are depicted in Fig.~\ref{fig:normal2}, conforming to the predetermined inter-vehicle spacing distance $(d_{1,2}=4m, d_{3,2}=10m, d_{4,2}=20m)$. 

The switching signal comprises two subsystems: the first is illustrated in Fig.~\ref{graph}(b), while the second is indicated in Fig.~\ref{graph}(c), revealing the process of retrieving the attacked agent. Therefore, by switching between two subsystems, namely Fig.~\ref{graph}(b) and Fig.~\ref{graph}(c), facilitated by the switching system, the depicted results are attained. 

 
	

\section{CONCLUSION} \label{sec:conclusion}
In this paper, a feasible vehicular resilient control strategy was presented to keep all members of the platoon safe when the leader of the network is under DoS attack. Once the leader is detected as an attacked vehicle, it will change its role to a follower in a platoon. It is noteworthy that after changing the role of the previous leader to a follower, it would be isolated from the rest of the vehicles to prevent it from sharing its contaminated data with other members of the platoon. By applying a proposed resilient controller, the attacked vehicle will be retrieved and returned to the leader-follower consensus and follow the new leader of the platoon. 

In future work, we plan to expand the current study to include platoons with heterogeneous or uncertain dynamics. Furthermore, machine learning algorithms can serve as a practical tool for self-driving vehicles to learn appropriate responses to DoS attacks.





\bibliography{main}

\begin{thebibliography}{10}

\bibitem{1}
J.~Shao, W.~X. Zheng, T.-Z. Huang, and A.~N. Bishop, ``On leader--follower consensus with switching topologies: An analysis inspired by pigeon hierarchies,'' {\em IEEE Transactions on Automatic Control}, vol.~63, no.~10, pp.~3588--3593, 2018.

\bibitem{2}
Y.~Zheng, S.~E. Li, K.~Li, and W.~Ren, ``Platooning of connected vehicles with undirected topologies: Robustness analysis and distributed h-infinity controller synthesis,'' {\em IEEE Transactions on Intelligent Transportation Systems}, vol.~19, no.~5, pp.~1353--1364, 2017.

\bibitem{3++}
A.~Buzachis, A.~Celesti, A.~Galletta, M.~Fazio, and M.~Villari, ``A secure and dependable multi-agent autonomous intersection management (ma-aim) system leveraging blockchain facilities,'' in {\em 2018 IEEE/ACM International Conference on Utility and Cloud Computing Companion (UCC Companion)}, pp.~226--231, IEEE, 2018.

\bibitem{3++1}
M.~Kargar, T.~Sardarmehni, and X.~Song, ``Optimal powertrain energy management for autonomous hybrid electric vehicles with flexible driveline power demand using approximate dynamic programming,'' {\em IEEE Transactions on Vehicular Technology}, vol.~71, no.~12, pp.~12564--12575, 2022.

\bibitem{4}
B.~Wang, J.~Wang, B.~Zhang, W.~Chen, and Z.~Zhang, ``Leader--follower consensus of multivehicle wirelessly networked uncertain systems subject to nonlinear dynamics and actuator fault,'' {\em IEEE Transactions on Automation Science and Engineering}, vol.~15, no.~2, pp.~492--505, 2017.

\bibitem{5}
M.~Kargar, C.~Zhang, and X.~Song, ``Integrated optimization of powertrain energy management and vehicle motion control for autonomous hybrid electric vehicles,'' {\em IEEE Transactions on Vehicular Technology}, 2023.

\bibitem{6}
D.~Zhang, G.~Feng, Y.~Shi, and D.~Srinivasan, ``Physical safety and cyber security analysis of multi-agent systems: A survey of recent advances,'' {\em IEEE/CAA Journal of Automatica Sinica}, vol.~8, no.~2, pp.~319--333, 2021.

\bibitem{pasqualetti2013attack}
F.~Pasqualetti, F.~D{\"o}rfler, and F.~Bullo, ``Attack detection and identification in cyber-physical systems,'' {\em IEEE transactions on automatic control}, vol.~58, no.~11, pp.~2715--2729, 2013.

\bibitem{7}
C.~Deng and C.~Wen, ``Mas-based distributed resilient control for a class of cyber-physical systems with communication delays under dos attacks,'' {\em IEEE transactions on cybernetics}, vol.~51, no.~5, pp.~2347--2358, 2020.

\bibitem{256+1}
C.~Boateng, K.~Yang, S.~G.~A. Ghoreishi, J.~Jang, M.~T. Jan, J.~Conniff, B.~Furht, S.~Moshfeghi, D.~Newman, R.~Tappen, {\em et~al.}, ``Abnormal driving detection using gps data,'' in {\em 2023 IEEE 20th International Conference on Smart Communities: Improving Quality of Life using AI, Robotics and IoT (HONET)}, pp.~1--6, IEEE, 2023.

\bibitem{9}
M.~Yesilbudak and I.~Colak, ``Main barriers and solution proposals for communication networks and information security in smart grids,'' in {\em 2018 International Conference on Smart Grid (icSmartGrid)}, pp.~58--63, IEEE, 2018.

\bibitem{10}
L.~Fillatre, I.~Nikiforov, P.~Willett, {\em et~al.}, ``Security of scada systems against cyber--physical attacks,'' {\em IEEE Aerospace and Electronic Systems Magazine}, vol.~32, no.~5, pp.~28--45, 2017.

\bibitem{11}
F.~Liu, C.~Wang, and Q.~Geng, ``Observer-based mpc for ncs with actuator saturation and dos attacks via interval type-2 t--s fuzzy model,'' {\em IET Control Theory \& Applications}, vol.~14, no.~20, pp.~3537--3546, 2020.

\bibitem{12+1}
H.~Mokari, E.~Firouzmand, I.~Sharifi, and A.~Doustmohammadi, ``Deception attack detection and resilient control in platoon of smart vehicles,'' in {\em 2022 30th International Conference on Electrical Engineering (ICEE)}, pp.~29--35, IEEE, 2022.

\bibitem{12}
T.~Zhang and Z.~Li, ``Resilient network-level design of leader-follower multi-agent systems against dos attacks,'' in {\em 2020 39th Chinese Control Conference (CCC)}, pp.~5122--5127, IEEE, 2020.

\bibitem{13}
Z.~A. Biron, S.~Dey, and P.~Pisu, ``Resilient control strategy under denial of service in connected vehicles,'' in {\em 2017 American Control Conference (ACC)}, pp.~4971--4976, IEEE, 2017.

\bibitem{15}
G.~Wen, C.~P. Chen, H.~Dou, H.~Yang, and C.~Liu, ``Formation control with obstacle avoidance of second-order multi-agent systems under directed communication topology,'' {\em Science China Information Sciences}, vol.~62, no.~9, pp.~1--14, 2019.

\bibitem{17}
Z.~Zhao and H.~Shi, ``Semi-global output consensus of multi-agent systems subject to actuator saturation: A low and high gain approach,'' in {\em 2019 Chinese Control Conference (CCC)}, pp.~6235--6240, IEEE, 2019.

\bibitem{18}
Y.~Jiang, Y.~Zhang, and S.~Wang, ``Distributed adaptive consensus control for networked robotic manipulators with time-varying delays under directed switching topologies,'' {\em Peer-to-Peer Networking and Applications}, vol.~12, no.~6, pp.~1705--1715, 2019.

\bibitem{19}
Y.~Cao, Y.~Sun, and X.-J. Xie, ``Explicit condition for consensus of third-order discrete-time multi-agent systems without accelerated velocity measurements,'' {\em Journal of The Franklin Institute}, vol.~354, no.~12, pp.~5056--5066, 2017.

\bibitem{20}
Q.~Shi, T.~Li, J.~Li, C.~P. Chen, Y.~Xiao, and Q.~Shan, ``Adaptive leader-following formation control with collision avoidance for a class of second-order nonlinear multi-agent systems,'' {\em Neurocomputing}, vol.~350, pp.~282--290, 2019.

\bibitem{23}
R.~Toyota and T.~Namerikawa, ``Formation control of multi-agent system considering obstacle avoidance,'' in {\em 2017 56th Annual Conference of the Society of Instrument and Control Engineers of Japan (SICE)}, pp.~446--451, IEEE, 2017.

\bibitem{23+1}
M.~T. Jan, C.~Garbin, J.~Ruetschi, O.~Marques, and H.~Kalva, ``Automated patient localization in challenging hospital environments,'' {\em Multimedia Tools and Applications}, pp.~1--19, 2024.

\bibitem{256}
H.~Mokari, E.~Firouzmand, I.~Sharifi, and A.~Doustmohammadi, ``Dos attack detection and resilient control in platoon of smart vehicles,'' in {\em 2021 9th RSI International Conference on Robotics and Mechatronics (ICRoM)}, pp.~144--150, IEEE, 2021.

\bibitem{3}
H.~Mokari, E.~Firouzmand, I.~Sharifi, and A.~Doustmohammadi, ``Resilient control strategy and attack detection on platooning of smart vehicles under dos attack,'' {\em ISA transactions}, vol.~144, pp.~51--60, 2024.

\bibitem{26}
D.~Zheng, H.~Zhang, and Q.~Zheng, ``Consensus analysis of multi-agent systems under switching topologies by a topology-dependent average dwell time approach,'' {\em IET Control Theory \& Applications}, vol.~11, no.~3, pp.~429--438, 2017.

\bibitem{278}
K.~Gu, J.~Chen, and V.~L. Kharitonov, {\em Stability of time-delay systems}.
\newblock Springer Science \& Business Media, 2003.

\end{thebibliography}
\bibliographystyle{ieeetr}


\end{document}